\documentclass[prc]{revtex4}

\usepackage{graphicx,amsmath}

\begin{document}

\title{The Quark-Gluon Plasma in a Finite Volume}

\author{Scott Pratt}
\email{pratts@pa.msu.edu}
\affiliation{Department of Physics and Astronomy, Michigan State
University, East Lansing Michigan, 48824}
\author{J\"org Ruppert}
\email{ruppert@th.physik.uni-frankfurt.de}
\affiliation{Institut f\"ur Theoretische Physik, J. W. Goethe-Universit\"at,
Robert-Mayer-Str. 8-10, D-60054 Frankfurt am Main, Germany}
\date{\today}

\begin{abstract}
\bigskip
The statistical mechanics of quarks and gluons are investigated within the
context of the canonical ensemble.  Recursive techniques are developed which
enforce the exact conservation of baryon number, total isospin, electric
charge, strangeness, and color. Bose and Fermi-Dirac statistics are also
accounted for to all orders. The energy, entropy and particle number densities
are shown to be significantly reduced for volumes less than 5 fm$^3$.
\end{abstract}

\maketitle

\section{Introduction}
\label{sec:intro} 

The study of the quark gluon plasma in the laboratory through hadronic
collisions represents one of the largest current initiatives in nuclear
physics. Although these volumes are manifestly finite, modeling of the
collisions seldom addresses the impact of the finiteness as most treatments are
based on the the grand canonical ensemble where the conservation constraints of
baryon number, strangeness, isospin and color are ignored. Indeed, one expects
that these constraints would become meaningless for larger volumes. The
effective volume for a $pp$ collision might be only a few cubic fm, while the
volume of a RHIC collision might be a few thousand cubic fm. But even in a RHIC
collision, the duration of the quark gluon phase might be only a few fm/c,
meaning that the effective volume for charge conservation may be a few dozen
cubic fm.

Recently, the effects of local charge conservation have gained more attention
due to the relationship of charge conservation to charge fluctuations and
charge balance
functions\cite{bassdanpratt,jeonkoch,asakawaprl,asakawa,gavinqm2002,gavinkapusta,pruneaugavinvoloshin,shuryakstephanov,stephanovqm2001}. These
observables are intimately related to the delay of hadronization expected
should a quark-gluon plasma be created. Recent papers have focused on the
importance of conserving electric charge, strangeness and baryon number, as
well as the dynamics of charge equilibration \cite{kochlin}. The statistical
mechanics of a finite-volume hadron gas have been studied in the context of
pion flavor distributions\cite{senisofluc}.

The approaches for solving the canonical ensemble can be divided into two
classes. In the first set of approaches, one can sum over all partition
functions in the grand canonical ensemble assuming the chemical potentials are
imaginary, $Z\sim {\rm Tr}e^{-\beta H +i\beta\mu Q}$. By integrating over all
$\mu$, the phase factor becomes a delta function which restricts the phase
space to states with $Q=0$ \cite{becattini,turko,redlichturko}. This approach
can even be applied to non-Abelian symmetries such as
SU(N)\cite{elzeplb,elzepra}. By associating $N-1$ charges, $Q_1\cdot Q_{N-1}$,
with the $N-1$-fold Cartan subgroup, one can restrict the ensemble to states
with specific charges $Q$, which can then be associated with the SU(3)
multiplet labels, $(p,q)$, by a transformation. These techniques work well for
simple systems, e.g. massless partons, where the integral over complex phases
can be be performed analytically. Multiple charges can be readily incorporated,
including conservation of overall momentum.

The second class of approaches for the canonical ensemble centers about
recursion relations which were proposed by Chase and
Mekjian\cite{mekjian,dasguptamekjian} for the study of nuclear
fragmentation. These methods have been extended to include quantum
statistics\cite{prattfermi} and non-additive charges\cite{senisofluc}, such as
angular momentum and isospin. In addition to partition functions, these methods
can also generate multiplicity
distributions\cite{prattdasgupta,senisofluc}. The advantage of these methods is
that, given the one-particle partition functions, they provide exact
answers. Since these methods involve sums, rather than integration over phases,
they provide robust answers for arbitrary energy levels. Furthermore,
symmetrization is included to all orders. If the number of particles to be
considered remains below 100, numerical calculations tend to take only seconds
or minutes at most.

We review recursive techniques for canonical ensembles and present extensions
to incorporate conservation of color in the next section. In the subsequent
section a simple example of a non-interacting quark-gluon gas is
explored. Canonical partition functions are generated which enforce
conservation of baryon number, strangeness and charge, while requiring the
system to be in an isosinglet and in a color singlet. Bose and Fermi statistics
are included to all order.  Energy, entropy, and particle number densities are
all shown to be significantly suppressed for small volumes, less than 5 fm$^3$.
Arguments are made that these effects are unlikely to be important in a
relativistic heavy ion collision, but might be quite important for $pp$
collisions.

\section{Recursive techniques for quantum colored particles}

For additive charges, canonical partition functions can be calculated in a
straight-forward manner. Neglecting symmetrization, conservation of a set of
charges $\vec{Q}$ can be enforced through the recursion relation,
\begin{equation}
Z_{A,\vec{Q}}(T)=\sum_k \frac{a_k\omega_k(T)}{A}
Z_{A-a_k,\vec{Q}-\vec{q}_k}(T).
\end{equation}
Here, $k$ refers to all particles species of charge $\vec{q}_k$ whose
single-particle partition functions are:
\begin{equation}
\omega_k = \sum_i d_i e^{-E_i/T},
\end{equation}
where $i$ denotes the individual energy levels of energy $E_i$ and degeneracy
$d_i$. The number $a_k$ must be positive for every species. For low energy
nuclear physics applications, $a_k$ could be the baryon number and $A$ would
represent one of the conserved charges. For a high-energy system where
anti-particles must be included and there are no positive-definite conserved
charges, $A$ can refer to the number of hadrons. This number is
positive-definite, but as the number of hadrons is not conserved, one must
consider all $A$ to obtain the partition function, $Z=\sum_A Z_A$.

Symmetrization can be incorporated by considering all permutations of identical
particles \cite{prattfermi,prattmultibose},
\begin{equation}
Z_{A,\vec{Q}}=\frac{1}{A!}
\sum_{j_1\cdots j_A,{\mathcal P}(i)}\langle j_1\cdots j_A|e^{-H/T}|
{\mathcal P}(j_1\cdots j_A)\rangle(-1)^{N_P({\mathcal P})},
\end{equation}
where a sum over all $A!$ permutations of the single particle levels accounts
for the symmetrization and $N_P({\mathcal P})$ counts the number of pair-wise
interchanges of fermions involved in the permutation ${\mathcal P}$.

By considering the cyclic permutation,
\begin{equation}
\begin{split}
C^{k,\ell}(T)\equiv& \sum_{j_1\cdots j_\ell}\langle j_1,j_2\cdots j_\ell|
e^{-H/T}|j_2,\cdots j_\ell,j_1\rangle\\
=&\sum_i e^{-\ell E_i/T}=\omega_k(T/\ell),
\end{split}
\end{equation}
one can derive a modified recursion relation,
\begin{equation}
Z_{A,\vec{Q}}(T)=\sum_k \sum_\ell\frac{a_kC^{k,\ell}(T)}{A}(\pm 1)^{\ell+1}
Z_{A-\ell a_k,\vec{Q-\ell q_k}}(T).
\end{equation}
By summing over all possible values of $\ell$, symmetrization is included to
all orders.

Before addressing SU(3) symmetries, we review the techniques used for SU(2) as
described in \cite{senisofluc}.  Extending the techniques described above to
incorporate an SU(2) symmetry such as isospin or angular momentum can be
accomplished in two ways. First, one can consider the projection $M$ as an
additive charge and solve for $Z_{M}(T)$ with the method outlined above. Given
the $(2I+1)$ degeneracy of all multiplets of size $I$, the partition function
for states with fixed $I$ is easily obtained from the partition function for
fixed $M$, $Z_I=Z_M-Z_{M+1}$. The second technique requires one to understand
the decomposition of the cycle diagrams into SU(2) multiplets. In terms of the
projection operator, ${\mathcal O}^I_M$, which projects states with total
isospin $I$ and projection $M$, one can define the isospin decomposition of the
cycle diagram.
\begin{equation}
\begin{split}
C^{\kappa,\ell}_{i}(T)
\equiv& \sum_{j_1\cdots j_\ell}\langle j_1,j_2\cdots j_\ell|
e^{-H/T}{\mathcal O}^i_m|j_2,\cdots j_\ell,j_1\rangle\\
=&\omega_\kappa(T/\ell)\chi^{\kappa,\ell}_{i},
\end{split}
\end{equation}
where $\chi^{\kappa,\ell}_i$ is the operator $C^{\ell,i}_k$ evaluated with
the particles being combined to one quantum state of zero energy. The subscript
$\kappa$ refers to a specific multiplet, not the individual states. For
instance, $\kappa$ might refer to pions and $\omega_\kappa(T)$ would be the
partition function for a given species of pion. The details about the energy
levels are absorbed in $\omega_{\kappa}$ while $\chi$ is independent of the
energy levels, but does depend on the isospin of the particular species,
$i_{\kappa}$.

Once $\chi^{\kappa,\ell}_i$ is known, one can calculate the partition function
for fixed $I$ recursively.
\begin{equation}
\label{eq:recursivesu2}
Z_{A,\vec{Q},I}(T)=\sum_{\kappa} \sum_\ell
\frac{a_k\omega_k(T/\ell)}{A}(\pm 1)^{\ell+1}
Z_{A-\ell a_k,\vec{Q}-\ell \vec{q}_k,I^\prime}(T)\chi^{\kappa,\ell}_i
{\mathcal N}(I^\prime,i;I).
\end{equation}
One could have added indices $M$, $M'$ and $m$ onto $Z_{A,I}$, $Z_{A',I'}$ and
$\chi^{\kappa,\ell}_i$ respectively. However, as each of these functions
involve a trace over all states, and the Hamiltonian has no dependence on the
isospin projection, none of these quantities has any dependence on the
projection and the indices can be suppressed. The function ${\mathcal
N}(I^\prime,i,I)$ counts the number of multiplets of size $I$ that one obtains
by coupling $I^\prime$ with $i$. For SU(2), these obey a simple form.
\begin{equation}
\begin{split}
{\mathcal N}(I^\prime,i;I)=&\sum_{M',m}\langle I,M|I',M',i,m\rangle^2\\
=&\left\{
\begin{array}{c c}
1,&|I^\prime-i|\le I\le I^\prime +i\\
0,&{\rm otherwise}
\end{array}
\right.
\end{split}
\end{equation}
Before performing the recursive calculations described in
Eq. (\ref{eq:recursivesu2}), one needs a simple expression for
$\chi^{\kappa,\ell}_i$. This can be accomplished by realizing that by summing
over $i$ one obtains the projection of states with fixed $m$.
\begin{equation}
\sum_i\chi^{\kappa,\ell}_{i,m}=\sum_{j_1\cdots j_\ell}
\langle j_1,j_2\cdots j_\ell|{\mathcal O}_m|j_2\cdots j_\ell,j_1\rangle.
\end{equation}
Since the states $|j_1\cdots j_\ell\rangle$ are eigenstates of ${\mathcal O}_m$,
the bra and ket must be identical, i.e., all particles must be identical and
have the same isospin projection.
\begin{equation}
\label{eq:isosum}
\sum_i\chi^{\ell,i}_{\kappa,m}=\left\{
\begin{array}{cc}
1,&m=0,\pm \ell,\pm 2\ell,\cdots \pm\ell i_\kappa\\
0,& {\rm otherwise}
\end{array}
\right.
\end{equation}
Since $\chi^{\kappa,\ell}_{i,m}$ is independent of $m$ when $m$ is a member of
the isomultiplet $i$, the sum over $i$ in Eq. (\ref{eq:isosum}) can be limited
to $i\ge m$. It is then straight-forward to see that
\begin{equation}
\chi^{\kappa,\ell}_i=\left\{
\begin{array}{rc}
1,&i=0,\ell,2\ell,\cdots \ell i_\kappa\\
-1,&i=\ell-1,2\ell-1,\cdots \ell i_\kappa-1\\
0,&{\rm otherwise}
\end{array}
\right.
\end{equation}
Here, the index $m$ is suppressed since $\chi$ is independent of $m$ as long as
$m$ is a member of the isomultiplet $i$. For the $i_\kappa=1$ case, e.g., pions
or $\rho$ mesons, $\chi^{\ell}_i=1$ for $i=\ell$ and $i=0$, and
$\chi^{\ell}_i=-1$ for $i=\ell-1$.

Restricting SU(3) color can be accomplished with the same procedure. Following
the same steps as the SU(2) case, one can derive analogous expressions,
\begin{equation}
\label{eq:recursivesu3}
Z_{A,\vec{Q},(P,Q)}(T)=\sum_{\kappa,\ell,(P',Q),(p,q)}
\frac{a_\kappa\omega_\kappa(T/\ell)}{A}(\pm 1)^{\ell+1}
Z_{A-\ell a_\kappa,\vec{Q}-\ell \vec{q}_\kappa,(P',Q')}(T)
\chi^{\kappa,\ell}_{(p,q)}
{\mathcal N}((P',Q'),(p,q);(P,Q)).
\end{equation}
Here, $(P,Q)$ denotes an SU(3) multiplet, e.g., the gluon color octet is
represented by $(P=1,Q=1)$. For an explanation fo the notation see
\cite{youngtableauxref}. Again, the challenge in making
Eq. (\ref{eq:recursivesu3}) tractable is in finding expressions for
$\chi^{\kappa,\ell}_{(p,q)}$ and ${\mathcal N}((P',Q'),(p,q);(P,Q))$. Coupling
SU(3) multiplets, i.e., finding expressions for ${\mathcal N}$, follows rules based
on manipulating Young tableaus \cite{youngtableauxref}. For the calculations in
this paper, these rules were programmed numerically.

Finding an expression for $\chi^{\kappa,\ell}_{(p,q)}$ requires performing a
color decomposition of the cycle diagrams and can be done analogously as was
shown above for SU(2). First, one must project states of a given $\mu$, where
$\mu$ represents the eigenvalues of the projection operators in SU(3), e.g.,
hyper-charge and $I_3$ for SU(3) flavor. As with the SU(2) example above, this
projection is realized by summing $\chi^{\kappa,\ell}_{(p,q),\mu}$ over all
$(p,q)$ multiplets which include the projection $\mu$.
\begin{equation}
\sum_{(p,q)}\chi^{\kappa,\ell}_{(p,q),\mu}=\sum_{j_1\cdots j_\ell}
\langle j_1,j_2\cdots j_\ell|{\mathcal O}_\mu|j_2\cdots j_\ell,j_1\rangle.
\end{equation}
Again, the $\mu$ dependence in $\chi$ represents the known degeneracy of $\mu$
within a given $(p,q)$ multiplet.

\begin{figure}
\centerline{\includegraphics[width=0.6\textwidth]{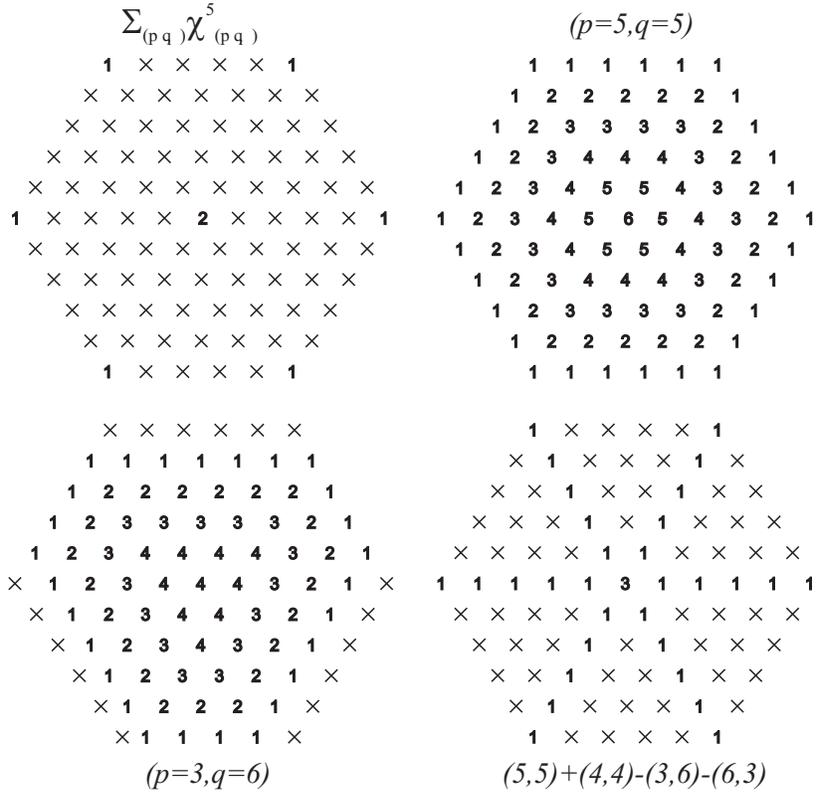}}
\caption{\label{fig:bosesu3} Upper-left: A diagrammatic
representation of the cyclic trace of five gluons in a single
quantum level. Upper-right and lower-left: A diagrammatic
representation of the $(5,5)$ and $(6,3)$ states. The integers
refer to the degeneracy of states with a given projection $\mu$.
Lower-right: By combining the four multiplets as shown, one
eliminates all states $\mu$ except those along diagonals defined
by the upper-right diagram. By taking the difference of these
states with the analogous combination for $\ell=4$, and adding two
color singlets, one obtains the desired combination of states in
the upper-left diagram.}
\end{figure}

At this point, we proceed by considering gluons as an example, and drop the
index $\kappa$. For $\ell$ gluons in a given quantum state the trace of
$\langle j_1,j_2\cdots j_\ell|{\mathcal O}_\mu|j_2\cdots j_\ell,j_1\rangle$ will be
zero unless $\mu$ corresponds to $\ell$ identical gluons. The upper-left panel
of Fig. \ref{fig:bosesu3} displays the projections $\mu$ that result from this
trace applied for five gluons. The upper-right and lower-left panels display
the $(\ell,\ell)$ and $(\ell-2,\ell+1)$ multiplets. By careful inspection, one
can see that the states along the diagonal, as shown in the lower-right panel
of Fig. \ref{fig:bosesu3}, are represented by the combination,
$(\ell,\ell)-(\ell+1,\ell-2)-(\ell-2,\ell+1)+(\ell-1,\ell-1)$.  By subtracting
the same object evaluated with $\ell'=\ell-1$ and adding two color singlets,
one can see that
\begin{equation}
\chi^{{\rm gluons},\ell}_{(p,q)}=
\left\{\begin{array}{rl}
1,&(p,q)=(\ell,\ell) ~{\rm or}~ (\ell-3,\ell) ~{\rm or}~ (\ell,\ell-3)\\
-1,&(p,q)=(\ell-2,\ell+1) ~{\rm or}~ (\ell+1,\ell-2) ~{\rm or}~
(\ell-2,\ell-2)\\
2,&(p=0,q=0)\\
0,&{\rm otherwise}
\end{array}
\right.
\end{equation}
Inserting this expression into Eq. (\ref{eq:recursivesu3}) allows one to
account for Bose effects in the gluonic partition function.

Quarks are represented by the multiplet $(p=1,q=0)$ while anti-quarks are
represented by the multiplet $(0,1)$. Following the same ideas as were
illustrated for the gluons, one can show
\begin{equation}
\chi^{{\rm quarks},\ell}_{(p,q)}=
\left\{\begin{array}{rl}
1,&(p,q)=(\ell,0) ~{\rm or}~ (\ell-3,0)\\
-1,&(p,q)=(\ell-2,1)\\
0,&{\rm otherwise}
\end{array}
\right.
\end{equation}

By combining the recursion relations for additive charges, which are used to
account for baryon and strangeness conservation, with the recursive techniques
for SU(2), which are used to enforce conservation of isospin, and the methods
for SU(3) which account for color, one can construct a recursive prescription
which account for all the charges in a parton gas. One might also choose to
write partition functions for subsystems, e.g., the strange quarks or the
gluons, then convolute the partition functions together to find the partition
function of the combined system. For instance, partitions for the two
subsystems $a$ and $b$ can be combined to find the combined partition function,
\begin{equation}
\label{eq:convolution}
Z_{(P,Q)}(T)=\sum_{(p_a,q_a),(p_b,q_b)} Z^a_{(p_a,q_a)}(T)Z^b_{(p_b,q_b)}(T)
{\mathcal N}((p_a,q_a),(p_b,q_b);(P,Q)).
\end{equation}

\section{Results and conclusions}
\label{sec:results}
The methods described in the last section were applied to the example of a
parton gas in a volume $V$ at a temperature, $T=250$ MeV. The gluons,
up-quarks and down-quarks were assumed to be massless, while the strange quark
was assumed to have mass of 150 MeV. The single-particle partition functions
were calculated by integrating over the momenta,
\begin{equation}
\begin{split}
\omega(T)=&\frac{2V}{(2\pi)^3}\int d^3p e^{-\beta\sqrt{p^2+m^2}}\\
=&\frac{mV}{2\beta^2\pi^2}\left(\beta mK_0(\beta m)+2K_1(\beta m)\right).
\end{split}
\end{equation}
The two helicities were accounted for by the factor of two preceding the
expression. One could easily account for shell effects by replacing the
integral over momentum with a discrete sum. Although discrete states are more
consistent given the Bose and Fermi effects described above, the continuous
form for $\omega$ is used here so that charge conservation effects can be
viewed separately.

The calculations were performed according to the following prescription:
\begin{enumerate}
\item The gluon partition function, $Z^{{\rm gluon}}_{(p,q)}$, was calculated
recursively by first calculating for all numbers of gluons $A$, then summing
over $A$.
\item The partition function for strange quarks, $Z^s_{A,(p,q)}$, was
generated. The partition function for anti-strange quarks,
$Z^{\bar{s}}_{A,(p,q)}$, was then generated by switching $(p,q)$ with
$(q,p)$. The partition function of the strange/anti-strange quark system was
found by summing over all partition functions with equal numbers of $s$ and
$\bar{s}$ quarks.
\begin{equation}
Z^{s\bar{s}}_{(p,q)}(T)=\sum_{A,(p_a,q_q)(p_b,q_a)}
Z^s_{A,(p_a,q_a)}(T)
Z^{\bar{s}}_{A,(p_b,q_b)}(T){\mathcal N}((p_a,q_a),(p_b,q_b);(p,q)).
\end{equation}
\item The partition function for up and down quarks were calculated separately.
In terms of the isospin projection, $m$, and the net number of up/down quarks,
$A$, the partition function for the up and down quarks is:
\begin{equation}
Z^{ud}_{A,m,(p,q)}(T)=
\sum_{(p_a,q_a)(p_b,q_b)}Z^{u}_{m+A/2,(p_a,q_a)}(T)
Z^{d}_{-m+A/2,(p_b,q_b)}(T){\mathcal N}((p_a,q_a),(p_b,q_b);(p,q)).
\end{equation}
\item The partition function for $\bar{u}\bar{d}$ quarks is calculated by
interchanging $p$ and $q$. In the same manner as the $s\bar{s}$ quarks were
convoluted, the $ud$ and $\bar{u}\bar{d}$ partition functions were convoluted
to find the partition function for $u$, $d$, $\bar{u}$ and $\bar{d}$ quarks
with the constraint of zero baryon number, $Z^{ud\bar{u}\bar{d}}_{M,(p,q)}$,
where $M$ is the isospin projection.
\item The partition function for the $ud\bar{u}\bar{d}$ system constrained to
  an isosinglet is calculated by taking the difference of the $M=0$ and $M=1$
  partition functions.
\item Using Eq. (\ref{eq:convolution}), the partition function for the
  $uds\bar{u}\bar{d}\bar{s}$ system were generated. This was then convoluted
  with the partition function for gluons to find the partition function for the
  entire system.
\end{enumerate}

Energy densities can be calculated with the well-known formula, $\langle
E\rangle= (-\partial/\partial\beta) \ln(Z)$. This requires performing the
calculation at two adjacent temperatures, doubling the CPU time of calculating
the partition function. Thus, an alternative approach was developed where
recursion relations were derived for $(\partial/\partial\beta) Z$. Recursion
relations were also generated for the trace of $Ne^{-\beta H}$, where $N$ could
be one of several number operators, $N_{\rm gluons}$, $N_{s}$, $N_u$ or
$N_d$. The recursion relations for these functions were calculated
simultaneously with the recursion relations for the partition functions with
little penalty in CPU time, because all calculations involve the same
calculations of ${\mathcal N}((p,q),(P',Q');(P,Q))$.

\begin{figure}
\centerline{\includegraphics[width=0.4\textwidth]{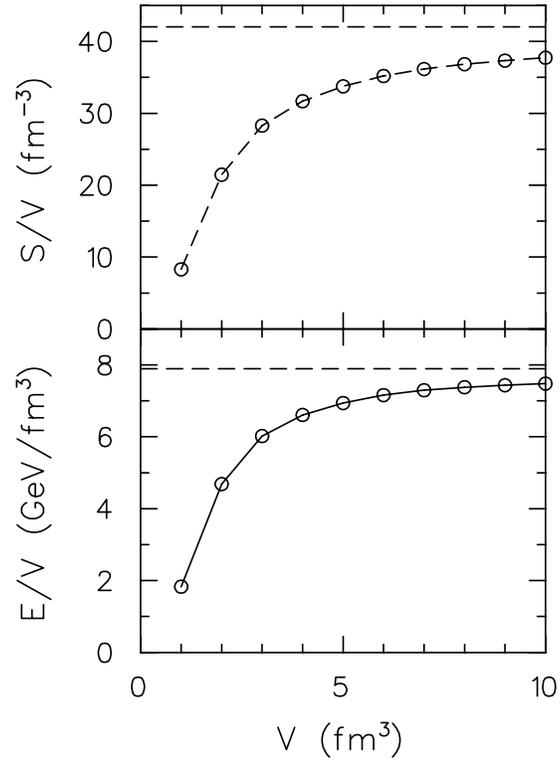}}
\caption{\label{fig:se}
The energy and entropy densities are shown for a parton system at a temperature
of 250 MeV as a function of the volume. For small volumes, conservation of
baryon number, isospin, strangeness and color restrict the phase space and
significantly lower the entropy and energy density. For larger volumes, the
results approach the grand canonical limit (dashed lines).}
\end{figure}

\begin{figure}
\centerline{\includegraphics[width=0.4\textwidth]{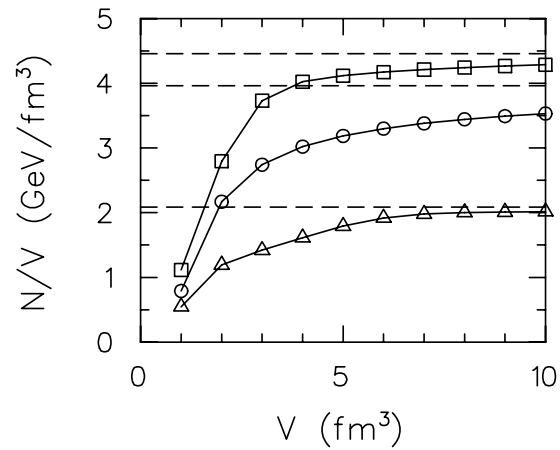}}
\caption{\label{fig:chemistry} Densities of gluons (circles), $s$ and $\bar{s}$
quarks (triangles), and $u,d,\bar{u}$ and $\bar{d}$ quarks (squares) are
suppressed for small volumes due to charge and color conservation. The grand
canonical limits are represented by dashed lines.}
\end{figure}

Figure \ref{fig:se} shows the energy and entropy densities as a function of the
volume. Both the entropy, $S=\ln(Z)+\beta\langle E\rangle$, and the energy are
lowered by the reduction in the available states due to the conservation
constraints. The grand canonical limit, represented by the dashed lines, is
approached at large volumes. As shown in \cite{elzeplb}, restricting the matter
to a color singlet reduces the partition function by a factor which scales as
$V^{-4}$. The other charge conservation constraints result in additional
reductions to the quark sectors. A reduction of $V^{-4}$ to the partition
functions corresponds to a reduction of $-4\ln(V)$ to the entropy and energy
densities. Since the bulk contributions scale as $V$ the logarithmic
conservation penalties become irrelevant at large volumes.  The chemical
decomposition of the various species behave similarly as shown in
Fig. \ref{fig:chemistry}. The color penalty for gluons is more severe than the
color penalties for quarks since gluons effectively have larger color charges
and are less likely to couple to a singlet.

From Fig.s \ref{fig:se} and \ref{fig:chemistry}, it is clear that conservation
rules are important for volumes of 5 fm$^3$ or less. It is therefore important
to take such considerations into account in $pp$ or $pA$ collisions. Heavy ion
collisions at RHIC occupy thousands of cubic fm. However, the local nature of
charge conservation results in an effective volume which is determined by the
initial conditions, and the diffusion of the various charges. Given that the
color of a projectile nucleon is spread over several units of rapidity by the
initial stopping process, and given the spread of charge over a few fm in the
transverse direction, it would be difficult to make a case that the effective
volumes are less than a few dozen cubic fm. It should be emphasized that the
determining factor for the importance of conservation constraints is the number
of partons in the effective volume. For massless partons, the density of
partons is expected to exceed 10 fm$^{-3}$ in a during the first one or two
fm/c of a RHIC collision. If the degrees of freedom are restricted by another
means, e.g., a large effective mass for partons, the relative penalty for color
conservation would increase.

\begin{acknowledgments}
This work was supported by the National Science Foundation, Grant No.
PHY-00-70818.  J.R. acknowledges support by the Studienstiftung des Deutschen
Volkes (German National Merit Foundation).  The authors wish to thank
D.D. Dietrich for sharing his profound understanding of SU(3).
\end{acknowledgments}

\end{document}